\documentclass[prl,aps,twocolumn,floats,superscriptaddress,longbibliography]{revtex4-1}
\usepackage{soul}
\usepackage{xcolor}
\usepackage{multirow}
\usepackage{amsmath}
\usepackage{graphicx}
\graphicspath{{figs/}}
\usepackage{amssymb}
\usepackage{epsfig}
\UseRawInputEncoding
\usepackage[T1]{fontenc}
\usepackage{pdfrender}

\usepackage[colorlinks = true]{hyperref}

\newlength{\upit}\upit=0.1truein

\newcommand{\ltappr}{{{\lower4pt\hbox{$<$} } \atop \widetilde{ \ \ \ }}}
\newlength{\bxwidth}\bxwidth=1.5 truein

\newcommand{\gtappr}{{{\lower4pt\hbox{$>$} } \atop \widetilde{ \ \ \ }}}

\newcommand{\be}{\begin{equation}}
\newcommand{\ben}{\begin{equation*}}
\newcommand{\ee}{\end{equation}}
\newcommand{\bea}{\begin{eqnarray}}
\newcommand{\eea}{\end{eqnarray}}

\newcommand{\een}{\end{equation*}}

\newcommand{\bmx}{\begin{array}}
\newcommand{\emx}{\end{array}}
\newcommand{\bean}{\begin{eqnarray*}}
\newcommand{\eean}{\end{eqnarray*}}

\newcommand{\beq}{\begin{equation}}
\newcommand{\eeq}{\end{equation}}

\newcommand\ltdash{\raise-0.7pt\hbox{$\scriptscriptstyle |$}}

\newlength{\figwidth}
\newlength{\shift}
\begin{document}
\title{Dual view of the Z$_2$-Gauged XY Model in 3D}

\author{Piers Coleman}
\affiliation{
Center for Materials Theory, Department of Physics and Astronomy,
Rutgers University, 136 Frelinghuysen Rd., Piscataway, NJ 08854-8019, USA}
\affiliation{ Department of Physics, Royal Holloway, University
of London, Egham, Surrey TW20 0EX, UK.}
\author{Anatoly Kuklov}
\affiliation{Department of Physics and Astronomy, CSI, and the
Graduate Center of CUNY, New York, USA}
\author{Alexei Tsvelik}
\affiliation{ Division of Condensed Matter Physics and Materials Science,
Brookhaven National Laboratory, Upton, NY 11973-5000, USA}
\pacs{72.15.Qm, 73.23.-b, 73.63.Kv, 75.20.Hr}
\begin{abstract}
The $Z_2$ gauged 
XY  model  is of 
long-standing interest both in the context of nematic order, and the study  
of fractionalization and superconductivity. 
This paper presents heuristic arguments that no deconfinement of the XY field occurs in this model and presents results of a large-scale Monte Carlo simulations  on a cubic lattice which are consistent with this conclusion. The correlation radius determining the confinement is found to be growing rapidly as a function of the parameters in the phase featuring the nematic order. Thus, mesoscopic properties of the system can mimic deconfinement with high accuracy in some part of the phase diagram. 
\end{abstract}

%
\maketitle
%

{\it Introduction. }In this  paper  we present a large-scale numerical
study of the classical Z$_2$-gauged XY model on a cubic lattice. This
model, 
described by the  Hamiltonian 
\be
H= - \frac{J}{2}\sum_{\langle i,j\rangle } u_{ij}[\psi^{*}_{i}\psi_{j}+ \psi^{*}_{j}\psi_{i}] - \kappa\sum_P\prod_{\langle i,j\rangle \in P} u_{ij},
 \label{H}
\ee
with $J>0, \kappa >0$, is the simplest realization of the class of models characterized
by higher symmetry matter fields interacting via $Z_{2}$ gauge
fields. Here the XY matter field $\psi_j = \exp(i\phi_j)$ is defined
at sites $j$, while
the $Z_2$ gauge field $u_{ij} = \pm 1$ is defined on the links of the 3D cubic 
lattice; the summation $\sum_{\langle ij\rangle} ...$ is performed over the links between the nearest neighbor sites $i$ and $j$ with each link contributing to the sum only once. The summation $\sum_P$ runs  over the elementary square plaquettes, with a product of the link variables defined around each plaquette.


The Z$_2$XY model (\ref{H}) arises in several different contexts. Its O(3) version 
was originally proposed in Refs.\cite{Lammert,Lammert2} as a
description of nematic liquid crystals, in which the term $\kappa$
acts to suppress disclinations, while the term $J$ promotes nematic
order. Later, the model acquired a
new significance 
in the context of high-temperature
superconductivity, as the
statistical mechanics 
counterpart to a $2+1$ dimensional description of electrons
fractionalized into spinless, charge $e$ bosons, described by the $\psi $ field,
interacting with the emergent $Z_{2}$ gauge field of a 
spin liquid \cite{Senthilfisher2001}. The topological phase transitions
in this model have also been related to the small-to-large Fermi
surface transitions in strongly correlated
electron systems \cite{Sachdev2018}.
Most recently, the Z$_2$XY model (\ref{H}) arose in the context of Kondo lattice models
coupled to an underlying $Z_{2}$ spin liquid \cite{ColemanTsv,CPT}  as the Ginzburg-Landau description of an order-fractionalized state with a charge $e$ order parameter.

\begin{figure}[t]
\includegraphics[width= \columnwidth]{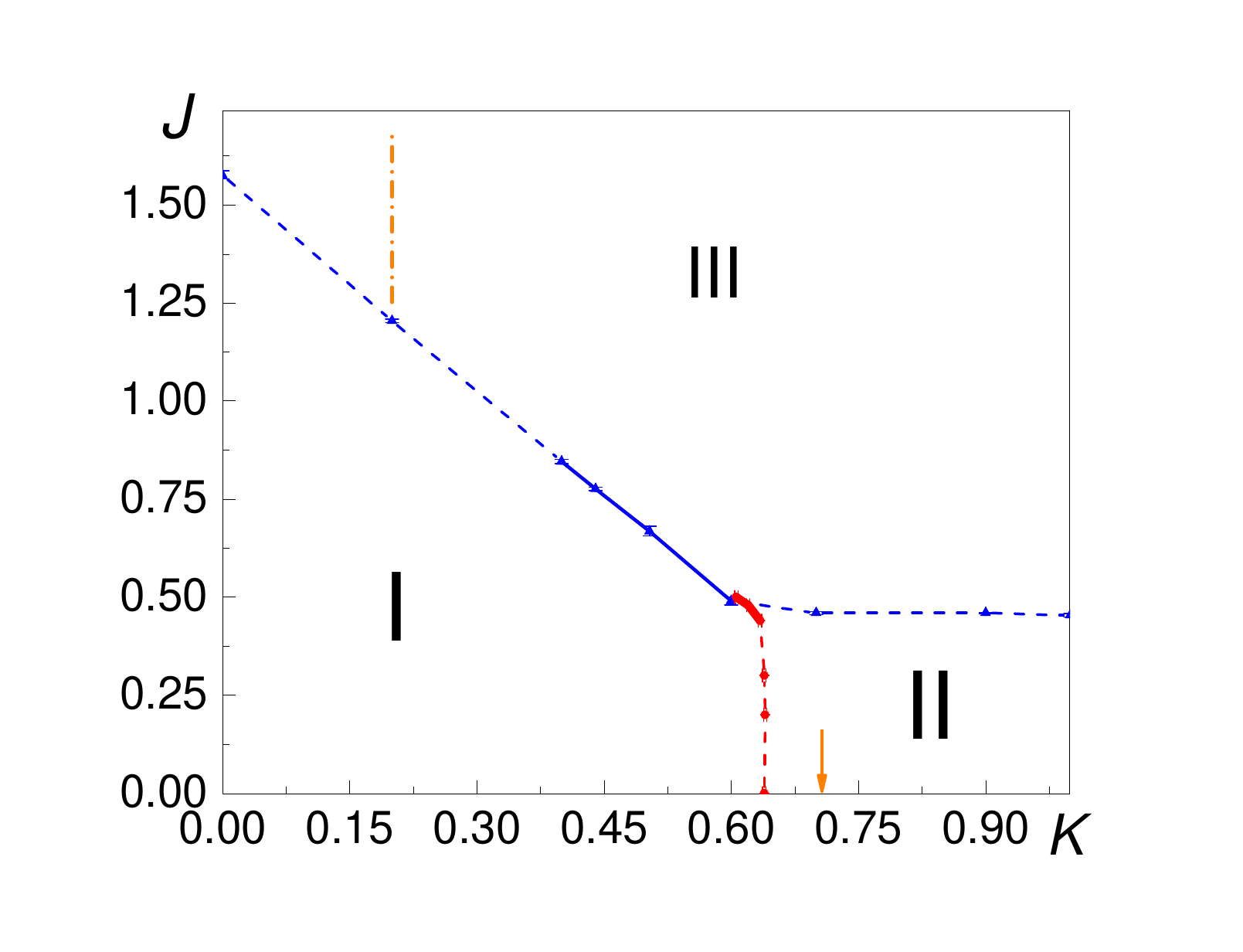}
\vspace{-0.2cm}
\caption{The phase diagram of model (\ref{H}) obtained in its dual form (\ref{Z3}), with $K = \tanh\kappa$. The dashed-dot line shows the range of $J$ where the data in Fig.~\ref{Fig:R12_K02} was collected at $K=0.2$. The arrow indicates the value $K=0.7$ for the data shown in Fig.~\ref{Fig:R2_L88}. The boundary between the phases I and II corresponds to the position of the heat capacity peak. Solid lines correspond to weakly 1st order transitions. }
\label{Fig:PD_fin}
\vspace{-0.5cm}
\end{figure}
\begin{figure}[htb]
\centerline {\includegraphics[width= \columnwidth]{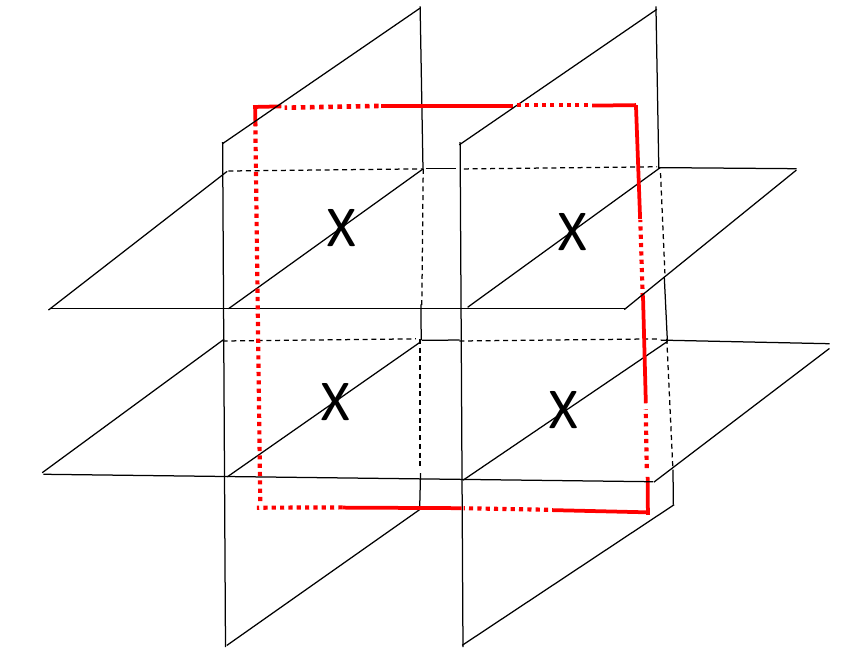}}
\vspace{-0.2cm}
\caption{An example of the vison loop (thick red line) piercing eight frustrated plaquettes. [The plaquette planes are visualized as semi-transparent]. The links marked by X are characterized by $u_{ij}=-1$. Accordingly, the energy of this vison loop is $E_{vis}=8\kappa$. The bonds with $u_{ij}=-1$ inside the vison loop induce a sign reversal in $\psi$, thus binding a vortex with a $\pi$ circulation to the vison. 
}
\label{Fig:vison}
\end{figure}

The model (\ref{H}) has three phases \cite{Scalapino_2002}:
a fully disordered phase (I) in which the matter field and the gauge sector are both disordered, a topologically ordered phase (II) in which the $Z_2$ gauge vortices (visons) are confined but the matter field is still disordered, and the superfluid phase where the matter field (III) condenses into an ordered state. In our simulations we have obtained the same phase diagram, Fig.~\ref{Fig:PD_fin}. While the phases I and II are well understood, the nature of the  superfluid phase III is still under discussion.  In Ref. \cite{Scalapino_2002} this phase was interpreted as a Bose condensate of the primary $\psi$-field, with an associated long-range order. This result, however, was obtained in Monte Carlo simulations of samples of linear sizes up to $L=8$. Larger sizes (up to $L=16$) have been simulated for the corresponding $Z_2$ gauged classical Heisenberg model, and the phase diagram of the same type as in Fig.~\ref{Fig:PD_fin}  was obtained where the phase III was found to be confined, corresponding to nematic or pair condensate in which the paired field $\langle \psi^2\rangle \neq 0$ is ordered, while the primary vector field remains disordered \cite{Lammert}. In a very recent study \cite{Vicari} it was found that, while no deconfinement occurs along the I-III line, the transition between II-III is deconfining. This finding is based on the results of the simulations of the linear sizes up to $L=40$ in a specific (stochastic)  gauge.

Discussions of  deconfinement traditionally focus on the behavior of the correlator $\langle\psi^*_i \psi_j\rangle$ in a specific gauge or in the situation when a string of the $Z_2$ field $u_{ij}$ connects the end points $i,j$ rendering the correlator gauge invariant.  In these cases the definition of the deconfinement is closely linked to the nature of the chosen gauge. 
As noted in Ref.\cite{Sachdev_book}, the deconfinement can be detected in the change of the power-law prefactor of the exponential decay of the gauge invariant correlator in the gapped phase.
Here we propose an approach based on a gauge independent definition of the deconfinement in the gappless phase using a duality transformation. 

{\it Vortex instability in Phase III. } Before discussing the duality we would like to give a gauge invariant argument that if the phase III were deconfining  at large $\kappa$, the nature of the topological excitations  would differ from those of a condensed $\psi =\exp(i \phi)$ field.  Using the observation \cite{Sachdev2018} that visons (Z$_2$ vortex-type excitations) bind to vortices (see Fig.~\ref{Fig:vison}), we note that a standard vortex with  2$\pi$ circulation of  $\phi$ in  phase III is unstable with respect to decay into two half vortices each carrying a vison  in its core,  provided the system size $L$ is large enough.
A configuration of a loop combining a $\pi$ vortex and a vison is sketched in Fig.~\ref{Fig:vison}. 
The key element of this configuration is that the energy of the vison per unit length of its core is $E_{vis} \approx \kappa $, while the energy per unit length of a $\pi$-vortex is $E_{q} \approx \pi q^2 J  \ln L$, where $q=1/2$ stands for half of the conventional $2\pi$ circulation. Comparing the energy $E_1$ of a standard vortex with $q=1$ with the energy $2E_{1/2} + 2E_{vis}= 2 [ \pi (1/2)^2 J\ln L + \kappa]$  of two half vortices each carrying a vison, we find that the difference  $  E_1 - 2(E_{1/2}+ E_{vis})=  (\pi/2) J \ln L - 2 \kappa$ becomes positive for $\ln L $ exceeding $ \approx 4\kappa /(\pi J)$, implying fission of the $2\pi$ vortex into two half vortex-vison bound states \cite{Sachdev2018}. From this argument, the nature of the phase III must be the same over the whole range $0\leq \kappa < \infty$, that is, characterized by nematic or $\psi^2$ order supporting $\pi$ vortices. In this paper we construct the dual representation of the model (\ref{Z3}) and formulate this 
argument from another perspective.

{\it Dual formulation. }
Here we derive the dual formulation of the model (\ref{H}) (for its Villain version see the supplemental).
The partition function $Z = \sum_{\{u_{ij}\}} \int D\phi\exp(-H)$ involves a product over
lattice links and plaquettes, 
\begin{widetext}

\be
Z = Z_0\sum_{\{u_{ij}\}} \int D\phi \prod_{<ij>} \exp\Big[Ju_{ij}\cos(\phi_i - \phi_j)\Big]\times
 \prod_p\Big[1 + K (u_1u_2u_3u_4)_p\Big], \label{Z}
\ee
\end{widetext}
where $K=\tanh\kappa$; $Z_0=[\cosh\kappa]^{N_p}$; $N_p=3L^3$ is the total number of plaquettes. Carrying out a high temperature expansion of $Z$ in powers of $J$ \cite{WA},
integrating over $\phi_i$ and summing over $u_{ij}$, we obtain 
\be
\frac{Z}{Z_0}=\sum_{\{M_p\}} K^{\sum_p M_p} \sum_{\{N^{\pm}_{ij}\}}\prod_{<ij>} \Delta_{ij} \frac{(J/2)^{N^{+}_{ij}+N^{-}_{ij}}}{N^{+}_{ij}! N^{-}_{ij}!}   ,  
\label{Z3}
\ee
where $N^{\pm}_{ij}=0,1,2, ...$ are powers of the Taylor expansion of $\exp( J \cos(\phi_i -\phi_j))$ in terms of $J \exp(\pm i(\phi_{i} -\phi_j))$ on each link; $M_p=0,1$  counts how many times a particular plaquette contributes to a term in the expansion of $Z$ in powers of $K$ in Eq.(\ref{Z}); $\Delta_{ij}=1-\mod(B_{ij},2) $ imposes the constraint that the number of gauge fields $u_{ij}$ on each link is even. We call this the link-parity constraint. Since each link is shared by 4 plaquettes, the power of $u_{ij}$ on a link $<ij>$ appearing inside the expansion is
\be
B_{ij}=N^+_{ij} + N^-_{ij} + M_{ij}=0,2,4,...   ,
\label{Bconst}
\ee
where $M_{ij}=0,1,2,3,4$  is the number of plaquettes adjacent to the link $\langle ij \rangle$ with $M_p=1$. 
Using the method \cite{Balian_Drouffe_Itzykson_1975} formulated in terms of the Ising spins located at vertices of the dual cubic lattice, $M_p$ of a  plaquette determines if the link of the dual lattice piercing this plaquette is frustrated ($M_p=1$) or not ($M_p=0$).  
Accordingly, in the case $J=0$ Eq.(\ref{Z3}) describes the Ising model, with the constraint (\ref{Bconst}) satisfied automatically.

The link variables $N^{\pm}_{ij}=0,1,2,3, ...$ satisfy the relation  $N^{+}_{ij}=N^{-}_{ji}$ and obey
Kirchhoff's current conservation law 
\be
\sum_{j=<i>} I_{ij} =0,  \quad I_{ij}= N^{+}_{ij} - N^{-}_{ij}
\label{Jconst}
\ee
with the summation performed over adjacent sites. This allows us to interpret the $I_{ij}=-I_{ji}$ as the conserved link currents flowing around closed oriented loops.  This expansion in terms of the current loops link variables closely follows Ref.\cite{WA}.
The current loops in the dual description are analogous to particle trajectories in a Feynman path integral description, an analogy that has been used in simulations of liquid Helium-4 \cite{Ceperley}. Our dual reformulation of the Z$_2$XY model is not only suited to worm-algorithm \cite{WA} computations, but it also enables us to discuss the phases in a gauge-invariant fashion. 

In the discussion that follows, we will refer to  the plaquette variables $M_p=0,1$ as  ``charges" and the link variables $I_{ij}$ as ``currents". These dual variables are $Z_2$  gauge invariant because the gauge fields $u_{ij}$ and $\phi_i$ have been integrated out. Presence of macroscopic loops of $I_{ij}$ is a necessary condition for the long range order in 3D and a finite superfluid stiffness, which in a 3D sample of linear size $L$ (chosen to be the same along each direction) is defined in terms of the windings numbers $W_\alpha$ of the loops as  
\cite{Ceperley}
\be
\rho_\alpha = \frac{1}{L} \langle W^2_\alpha \rangle, \, \, \, W_\alpha=\frac{1}{L} \sum_{\langle ij \rangle} I_{ij,\alpha}
\label{W}
\ee   
where $I_{ij,\alpha}$ are the link currents $I_{ij}$ along the $\alpha$-th direction. The quantity $\langle W^2_\alpha \rangle$ is scale invariant at the transition point of the standard XY model, and, thus, is used to identify the transition. The distinction between 
ordering in the $\psi$ or $\psi^2$ fields is characterized by the presence of large loops with either odd or even windings: deconfined order ($\psi$ condensed) involves odd-numbered loops, while confined order ($\psi^2$ condensed while $\psi$ is not)--- is characterized by even $W_\alpha$.  
This difference can be observed in the windings statistics ${\cal P}(W_\alpha)$. 
Thus, the deconfinement of the field $\psi$ requires that the configuration space of the loops and charges is characterized by a statistically significant presence of macroscopic loops with odd values of the windings in the limit $L\to \infty$. For $K=0$, only $M_p=0$ is possible in (\ref{Z3}), that is, the constraint (\ref{Bconst}) requires $I_{ij}=0,2,4,...$, which implies the confinement because no odd loops exist.

The development of order in the standard XY model (long range in 3D and algebraic in 2D) can be understood in terms of a proliferation of loops with $|I_{ij}|=1$, controlled by the free energy as a function of loop length $\sim L$. Indeed, its energy scales as $\sim -\ln(J/2)L $, while its entropy due to  shape fluctuations behaves as $\sim \ln(2d-1) L $. Thus, for $J > J_c \approx 1/(d-1/2) =2/5$ in 3D and $J_c\approx 2/3$ in 2D, loops proliferate and, accordingly, the $\psi$ order develops (cf. $J_c \approx 0.454$ in 3D \cite{Martin} and $J_c =2/\pi$ in 2D as the universal jump). 

In the Z$_{2}$XY model (\ref{Z3}) the free-energy argument requires a significant modification due to the link-parity constraint  (\ref{Bconst}). In the limiting  case where $J=0$, the bond constraint $B_{ij}= M_{ij}=0,2,4$ implies that the lowest energy non-trivial configuration of the dual variables in 3D is an  elementary cube formed of six plaquettes each carrying  charge $M_p=1$.  In the dual picture of the pure $Z_2$ gauge theory \cite{Balian_Drouffe_Itzykson_1975}, these elementary cubes map precisely onto the spin-flips of a 3D Ising model  with coupling constant $ - \frac{1}{2}\ln K$. 
Such a configuration automatically satisfies the constraint(\ref{Bconst}) with $B_{ij}=2$ on each of its edges. 
Rewriting (\ref{Z3}) as
$Z \sim \sum_{\{M_p\}}\exp(\ln K \sum_p M_p)$, one can interpret $\epsilon_1 = -\ln K$ as the excitation energy for the charges $M_p$. 
The total energy of such a cube is $-6\ln K$. As $K\to 0$ the density of the "charged" cubes drops rapidly to zero,  giving rise to a dilute gas of the cubes with the mean density  $m_p=\langle M_p\rangle \sim \exp(-6 \epsilon_1)\sim K^6 \to 0 $ (per plaquette).  This picture gives a dual view of the nature of  phase I, which in the direct variables corresponds to a soup of vison loops of arbitrary size.  In the variables of Ref.~\cite{Balian_Drouffe_Itzykson_1975} this phase corresponds to the Ising order of the dual spins. 

As $K$ approaches the boundary of 
  the phase II , $K=K_c \approx 0.639 \pm 0.003$ (cf. $K_c=\tanh(0.7613)\approx 0.6418$ from Ref.\cite{Balian_Drouffe_Itzykson_1975}) the cubes proliferate, eventually fusing into large-scale domain structures. [The supplemental gives  more detail]. Such correlations result in a 3D Ising transition characterized by the singularity in the specific heat. As $K$ approaches unity in phase II, simulations show that  the average density of the charges $m_p =\langle M_p\rangle$ tends to 1/2, corresponding to a disordered lattice of dual spins \cite{Balian_Drouffe_Itzykson_1975}  with an entropy of $\ln 2$ per link (that is, per  plaquette of the direct lattice).   
\begin{figure}[htb]
\centerline {\includegraphics[width=\columnwidth]{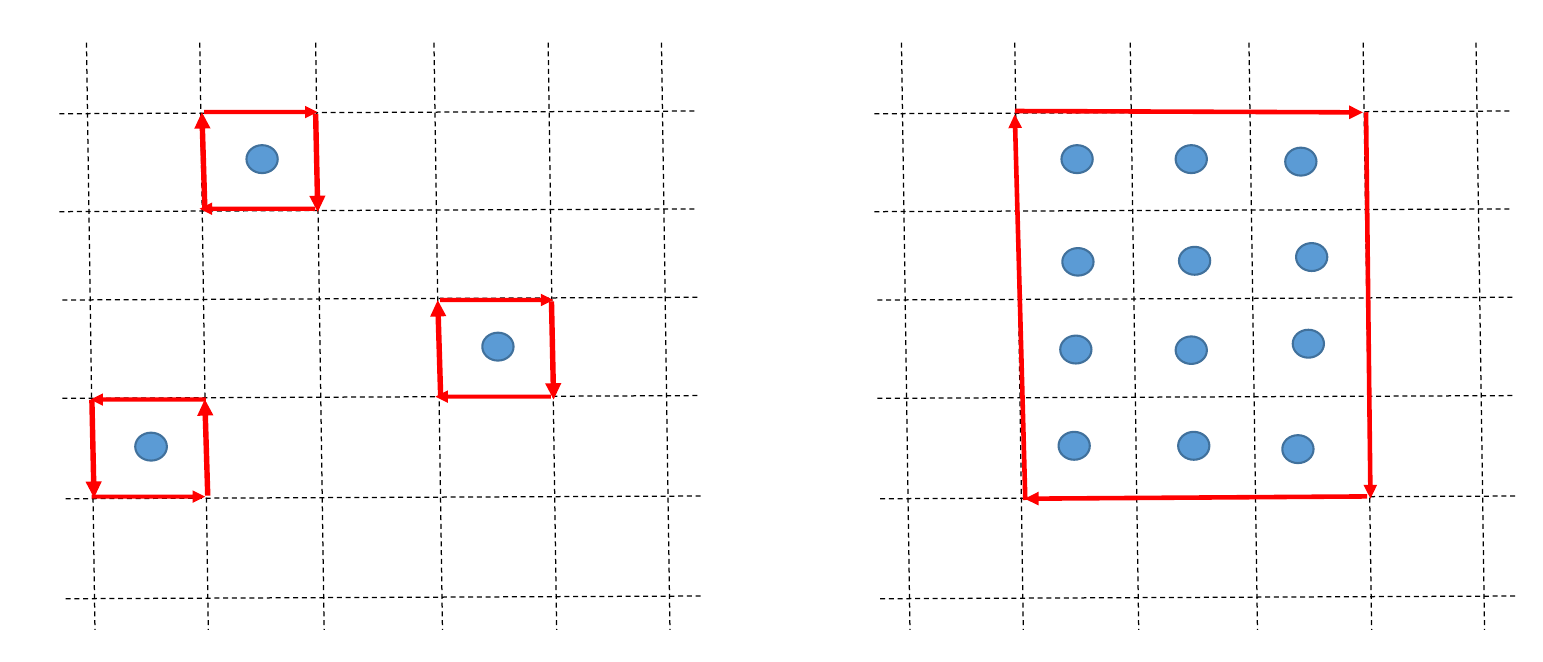}}
\vspace{-0.2cm}
\caption{Typical configurations of loops of dual odd link-currents with $|I_{ij}|=1$ (arrows) and the plaquette charges $M_p=1$ ( circles) satisfying the constraint (\ref{Bconst}). Left panel: a low density gas of the elementary loops with the charges. Right panel: a macroscopic loop with the membrane of the charges.  }
\label{Fig:Mgas2}
\end{figure}

The key question is what happens at finite $J$? [It is important to realize that for finite $J$ the mapping \cite{Balian_Drouffe_Itzykson_1975} looses its advantage because the constraint (\ref{Bconst}) cannot be satisfied automatically anymore]. For finite $J$ and small $K<<1$ the nature of the elementary excitation changes from the elementary cube to a plaquette surrounded by the elementary loop with $|I_{ij}|=1$ as sketched in the left panel in Fig.~\ref{Fig:Mgas2}.  The energy of  this object is $\epsilon_2 = - \ln K - 4 \ln J$. As $J$ increases, larger loops develop---as shown in the right panel of Fig.~\ref{Fig:Mgas2}. The energy of a loop of length $L$, together with its perimeter term   $E_P \sim  -L \ln J $, acquires the areal contribution $E_M \approx -\ln K L^2>0 $ from the membrane of the charges formed inside the loop perimeter. This gives the total energy as $E_M+E_P = -L  \ln J + L^2 |\ln K|$. Accordingly, the proliferation of the link-current loop with odd $I_{ij}$  beyond the size $L\sim \xi \approx (\ln J)/|\ln K| $ becomes statistically impossible even for $J>>1$.  This mechanism implies the confinement of the field $\psi$ because no odd windings $W$ can contribute to the stiffness (\ref{W}). Thus, the correlation length of the confinement becomes
\be
\xi \approx \ln J/ |\ln K|.
\label{xi}
\ee
It is important to note that this argument becomes asymptotically exact in the limit $K\to 0$ because the entropy of the membrane shape fluctuations cannot compete with the diverging membrane tension $\approx -\ln K$. 

We argue that Eq.(\ref{xi}) can also be applied to the transition from phase II to  phase III despite that the membrane tension decreases to  zero as $K\to 1$.
The reason for this is that the membrane formation can only lead to a decrease of the entropy from its maximal value $\ln 2$ per plaquette. This decrease is $\sim 1-K$ in the limit under consideration. Accordingly, the free energy of the membrane increases as $\delta F_M \sim 1-K \sim \exp(-2\kappa)$ leading to a divergence of the confinement length as $\xi \sim 1/(1-K) \sim \exp(2\kappa) \to \infty$. 
[A more accurate estimate of the numerical factor in the above relation can be found in the supplemental]. These arguments show that the phase III must be confining in both limits $\kappa \to 0$ and $\kappa \to \infty$ (with the exception $\kappa=\infty$ corresponding to the standard XY model).       

{\it Results of the simulations}.
\begin{figure}[thb]
\includegraphics[width=\columnwidth]{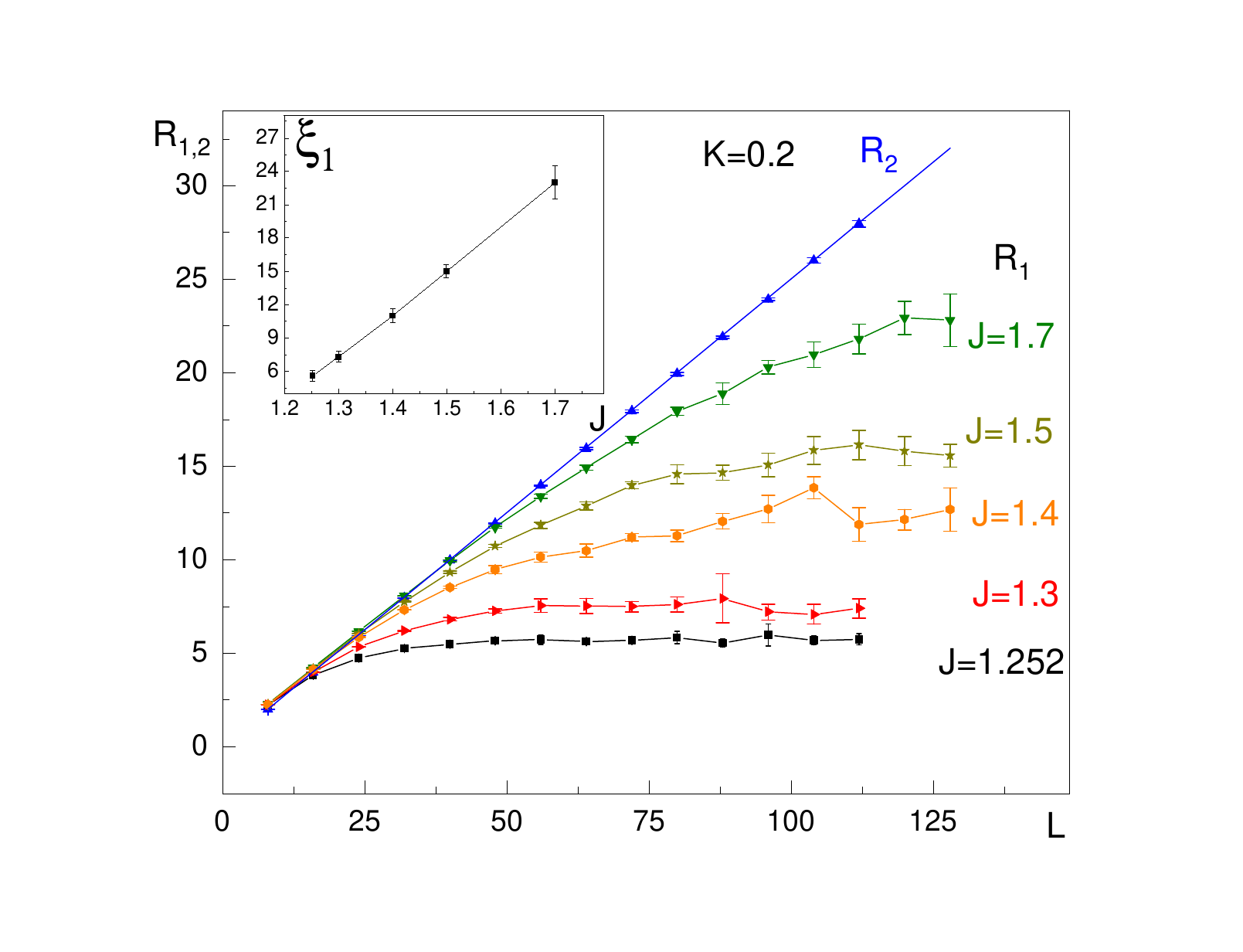} 
\vspace{-0.2cm}
\caption{Gyration radii  $R_{1,2}$ of $C_{1,2}$, respectively, plotted versus  system size $L$ at $K=0.2$,  for values of $J$ in the range 1.252-1.7 (see the vertical line in Fig.~\ref{Fig:PD_fin}). While $R_2$ follows the behavior $R_2 = L/4$ (in PBC) which corresponds to $C_2$=const, the $R_1$ curves saturate at $R_1\sim \xi (J)$ as $L\to \infty$. Inset: the corresponding correlation length $\xi(J)$ obtained as the mean of the sizes where $R_1$ saturated within the error bars, assuming that $C_1 \sim \exp(-|z|/\xi)$ for such sizes.  
}
\label{Fig:R12_K02}
\end{figure}
\begin{figure}[t]
\includegraphics[width=\columnwidth]{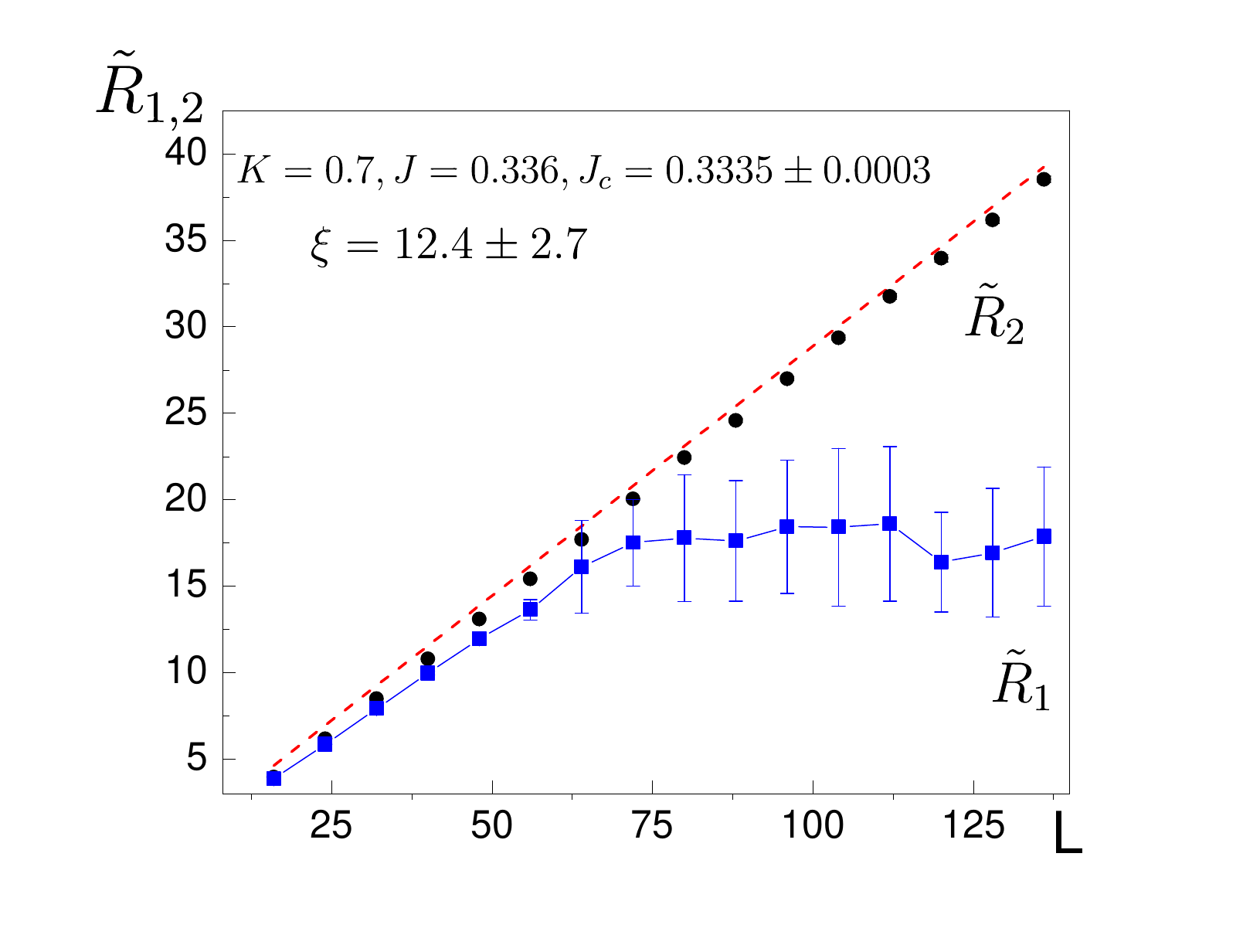} 
\vspace{-0.2cm}
\caption{ $ \tilde{R}_{1,2}$ vs system size $L$ for the J-current version of the model (\ref{Z}) for $K=0.7$ and $J=0.336$ (with the critical value $J_c=0.3335 \pm 0.0003$) in the phase III, where the stiffness $\rho_s \approx 0.038$. 
The dashed red line indicates the maximum possible value $L/(2\sqrt{3})$  of the radii for the uniform correlators (see the text).  The correlation radius $\xi$ for $C_1$ was determined as the mean of the values of $\tilde{R}_{1}/\sqrt{2}$ for $L > 90$, assuming that $C_1 \sim \exp(-|z|/\xi)$ for such sizes. }
\label{Fig:R2_L88}
\vspace{-0.5cm}
\end{figure}
Simulations using the Worm Algorithm (WA) \cite{WA} modified by the updates of the charges $M_p$ in order to satisfy the constraint (\ref{Bconst}) (see more details in the supplemental) 
have been conducted in a specific gauge $u_{ij}=1$ on all links along one lattice direction (say, $\hat{z}$). 
In this gauge there is a residual local symmetry when $\psi_i \to -\psi_i $ along the whole Z-line of sites at a given X and Y position of the site $i$. 
Accordingly, the correlation function $C_1(z)=\langle \psi^*_i \psi_j\rangle$ can be finite only if  X and Y coordinates of the points $i,j$ are the same.
The correlator $C_2=\langle \psi^{*2}_i \psi^2_j\rangle$ is not bound by this restriction because it is gauge invariant.  
To characterize the long-range correlations we have
calculated the radii of gyration using two definitions (for technical reasons to be explained later) as 
\begin{eqnarray}\label{gyr}
R_{1,2}= \frac{\sum_{x,y,z}|z| C_{1,2}}{\sum_{x,y,z}C_{1,2} }, \,\, \tilde{R}_{1,2}=\left[\frac{\sum_{x,y,z} z^{2}C_{1,2}}{\sum_{x,y,z}C_{1,2} }\right]^{\frac{1}{2}},
\end{eqnarray}
where $C_1$ depends only on the Z-coordinate while $C_2$ is a function of X,Y,Z.

Here we present results of the Monte Carlo simulations explicitly showing the confined nature of the phase III, first, at small but finite value of $K=0.2$, Fig.~\ref{Fig:R12_K02}, for the model (\ref{Z3}) slightly above the transition line between phases I and III. Inset to Fig.~\ref{Fig:R12_K02} indicates the dependence of the confining radius $\xi$ vs $J$. Fig.~\ref{Fig:R2_L88} shows the radii $\tilde{R}_{1,2}$ for the Villain version of the model (\ref{H}) at $K=0.7$ and $J=0.336$, that is, above the critical value $J_c=0.3335 \pm 0.0003$. 
It is worth noting that the limit $C_1 \sim \xi$ is reached starting from the sizes which are a factor 3-4 larger than $\xi$ itself.  In other words, observing the confinement at, say, $\kappa=1$ ($K\approx 0.76$) as compared to $\kappa=0.87$ ($K\approx 0.7)$ (see Fg.~\ref{Fig:R2_L88}) becomes computationally unrealistic.

{\it Discussion.} Our heuristic arguments indicate that the whole phase III is confined albeit with an exponentially diverging radius $\xi$ in the limit $\kappa \to \infty$. To falsify our claim we have looked for a possible transition line in the phase III featuring a singularity of the heat capacity which would occur, if the line between phases I and II extends to the phase III. 
The existence of such a line is a necessary condition for the deconfinement because of the broken Z$_2$ global symmetry associated with the deconfinement. No such a line has been found.   

There are a number of interesting caveats that accompany our results.  Although the $\psi$ field in  the $Z_2XY$ model remains confined in the ordered phase III, we note that its gauge-invariant correlation function does develop extremely long correlation lengths as long as $J$ exceeds its critical value $J_c$ for a given $K$. In the quantum analog of our model, this implies that the fractionalized field $\psi$ confines to form extremely diffuse bound-states. Here, it may be useful to make an analogy with quark confinement: in our model the fundamental field $\psi$ plays the role of a quark, while the composite $\psi^2$ is analogous to a meson. On scales beyond $\xi\sim e^{2\kappa}$  the $\psi $ field is confined, but on shorter scales,  the $\psi $ field is asymptotically free and the phase is indistinguishable from one in which $\psi$ is condensed.  If such a phenomenon develops in nature, it would imply the existence of superconducting phases  with charge $e$ order at short to intermediate distances. The strict symmetry description of the corresponding charge 2$e$ order parameter will then be formed from the product states of fractionalized charge e, spin 1/2 order parameters.  This would imply a much broader class of superconductivity than is currently known.

  We would also like to emphasize  that our main prediction following from the vortex-vison fission argument depends crucially on the neutrality of the $\psi$ field. It is interesting to consider the corresponding charged $Z_2$XY model,  in which the $\psi$ fields couple to both $Z_2$ and electromagnetic $U(1)$ fields, leading to a Z$_2\times U(1)$ gauge symmetry. In this case, we expect that  there {\sl will} be a deconfined phase IV, where vortices with  2$\pi$ 
  circulation in the phase of $\psi$ are stable. The key difference stems from the Meissner effect resulting in the screening out the log-tail at the London penetration length $\lambda_L$, so that for $\kappa \geq J \ln \lambda_L$ the $2\pi$-vortices will no longer split into half vortex-vison pairs. The corresponding boundary between the nematic phase at small $\kappa$ and the deconfined phase at large $\kappa$ is expected to exhibit a singularity in the heat capacity. We will address this model in future work.

\begin{acknowledgments}
		\textit{Acknowledgments:}
			We would like to thank Aaditya Panigrahi and Subir Sachdev for discussions related to this work.  This work was supported by Office of Basic Energy Sciences, Material
		Sciences and Engineering Division, U.S. Department of Energy (DOE)
		under Contracts No. DE-SC0012704 (AMT), by
National Science Foundation under the grant DMR-2335905 (AK) and DE-FG02-99ER45790 (PC ). Authors also acknowledge support from HPCC center of CUNY and the computational cluster Thalia, BNL. 
	\end{acknowledgments}
\section{Supplemental}	
Here we present the dual formulation based on the Villain approximation,  describe details of the algorithm and the nature of the configurations. 

\section{Duality based on the Villain approximation}
The partition function (\ref{Z}) in the Villain formulation reads
\begin{widetext}
\be
Z = Z_0\sum_{\{u_{ij}, m_{ij}\}} \int D\phi \prod_{<ij>} e^{-\frac{J}{2}(\phi_i - \phi_j + A_{ij} +2\pi m_{ij})^2} 
\prod_p\Big[1 + K (u_1u_2u_3u_4)_p\Big], \label{ZS}
\ee
\end{widetext}
where we used $u_{ij} \cos x =\cos(x + A_{ij} )$ with $A_{ij}=\pi (1-u_{ij})/2 $, given $u_{ij}=\pm 1$, and
in the Villain approximation $\exp(J\cos x) \to \sum_{m=0,\pm 1, \pm 2, ...} \exp( -\frac{J}{2} (x + 2\pi m)^2) $.
Then using the Poisson relation $\sum_{m=0,\pm 1, \pm 2, ...} f(m) = \sum_{I=0,\pm 1, \pm 2,...} \int dx f(x) \exp( 2\pi i I x)$, Eq.(\ref{ZS}) becomes after integrating out the phases as
\begin{widetext}
\be
Z = Z_0 \sum_{\{u_{ij}\}, \{I_{ij}\}} \prod_{<ij>} \Big[(u_{ij})^{I_{ij}}e^{-\frac{I_{ij}^2}{2J}}\Big] \times
 \prod_p\Big[1 + K \cdot (u_1u_2u_3u_4)_p\Big], \label{ZS}
\ee
\end{widetext}
with the constraint
\be
\sum_{j=<i>} I_{ij} =0, 
\label{JS}
\ee
which is the analog of Eq.(\ref{Jconst}). 
The factor $(u_{ij})^{I_{ij}}$ implies that for $K=0$ only even link currents $I_{ij}$ contribute to the partition function which means the confinement. This feature is fully in line with the discussion of this case for the action (\ref{H}) leading to  the action (\ref{Z3}).

Further summation over $u_{ij}$ gives
\be
Z=\sum_{\{M_p\},\{I_{ij}\}} K^{\sum_p M_p} \prod_{<ij>} \Delta_{ij}  \exp\Big[-\frac{I_{ij}^2}{2J}\Big]  ,  
\label{Z4}
\ee
with $\Delta_{ij} = 1- \mod(B_{ij},2)$ which selects
\be
B_{ij}=|I_{ij}| + M_{ij}=0,2,4,....
\label{BS}
\ee
where $M_{ij}$ has the meaning explained below Eq.(\ref{Bconst}) in the main text. 

 Simulations are performed in the dual representation (\ref{Z4}) using the WA \cite{WA} augmented by  additional updates of the plaquettes charges consistent with the parity requirements. The WA updates are of two types: i) with even link currents characterized by $|I_{ij}|=2,4,...$; ii) odd currents update $|I_{ij}|=1,3,...$ along $\hat{z}$-direction only.
 While the update i) does not change links parity in any direction, the update ii) retains the parity of the Z-links because of the gauge condition $u_{ij}=1$ along $\hat{z}$. 
The plaquette charges updates are also of two types: iii) swapping two values $M_p=0,1$ at all six sides of the elementary cube; iv) Swapping the values of $M_p$ in one plaquette and simultaneously creating (or removing) an elementary oriented loop (see the left panel in Fig.~\ref{Fig:Mgas2}) along this plaquette.     

\begin{figure}[thb]
\includegraphics[width=0.55\textwidth]{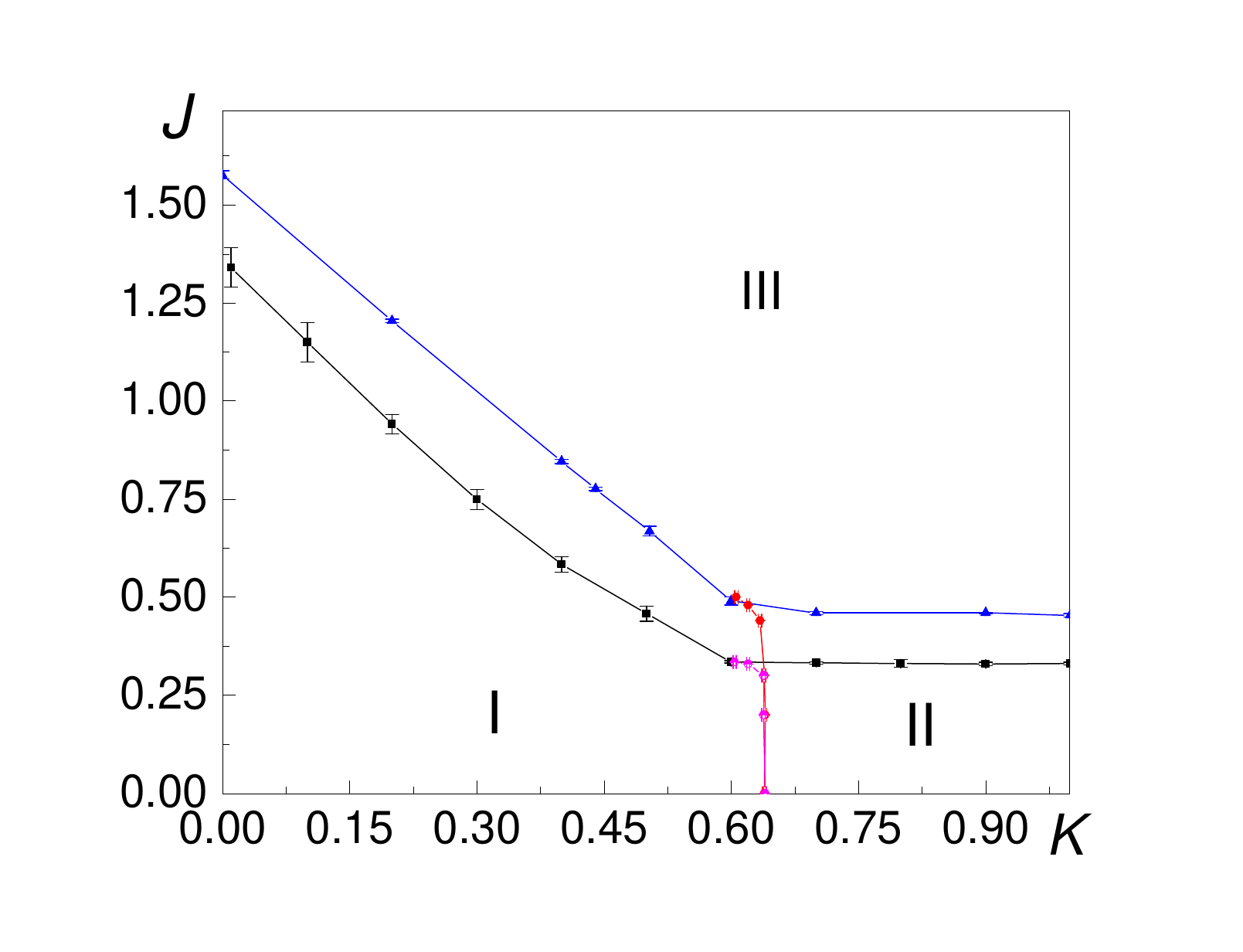} 
\vspace{-0.2cm}
\caption{The phase diagrams of the model (\ref{H})  (triangles connected by a solid blue line) and its Villain version (\ref{ZS}) (squares connected by the black solid line). The lines separating the phases I and II are overlapping for $J<0.3$: solid circles connected by the red line are for the model (\ref{H}) and the semi-open circles connected by the pink line are for the Villain version. 
}
\label{Fig:PD3}
\vspace{-0.5cm}
\end{figure}

The updates of the original model (\ref{Z3}) are similar to  those described above with only one difference that $I_{ij}$ is not the primary variable, and is determined as described in Eq.(\ref{Jconst}). Accordingly, additional updates on $N^{\pm}_{ij}$ are introduced. As it turns out, the convergence of the Villain version is better than that of the original one, especially, at larger $K$. This why we have used Villain version in simulations at $K=0.7$ (see Fig.~\ref{Fig:R2_L88}).
The phase diagrams of both models are very similar, with the phase boundary of the phase III in the 
Villain version being about 15-30 \% below that shown in the main text. As a comparison we present here both diagrams on a single graph, Fig.~\ref{Fig:PD3}.
\begin{figure}[t]
\includegraphics[width= \columnwidth]{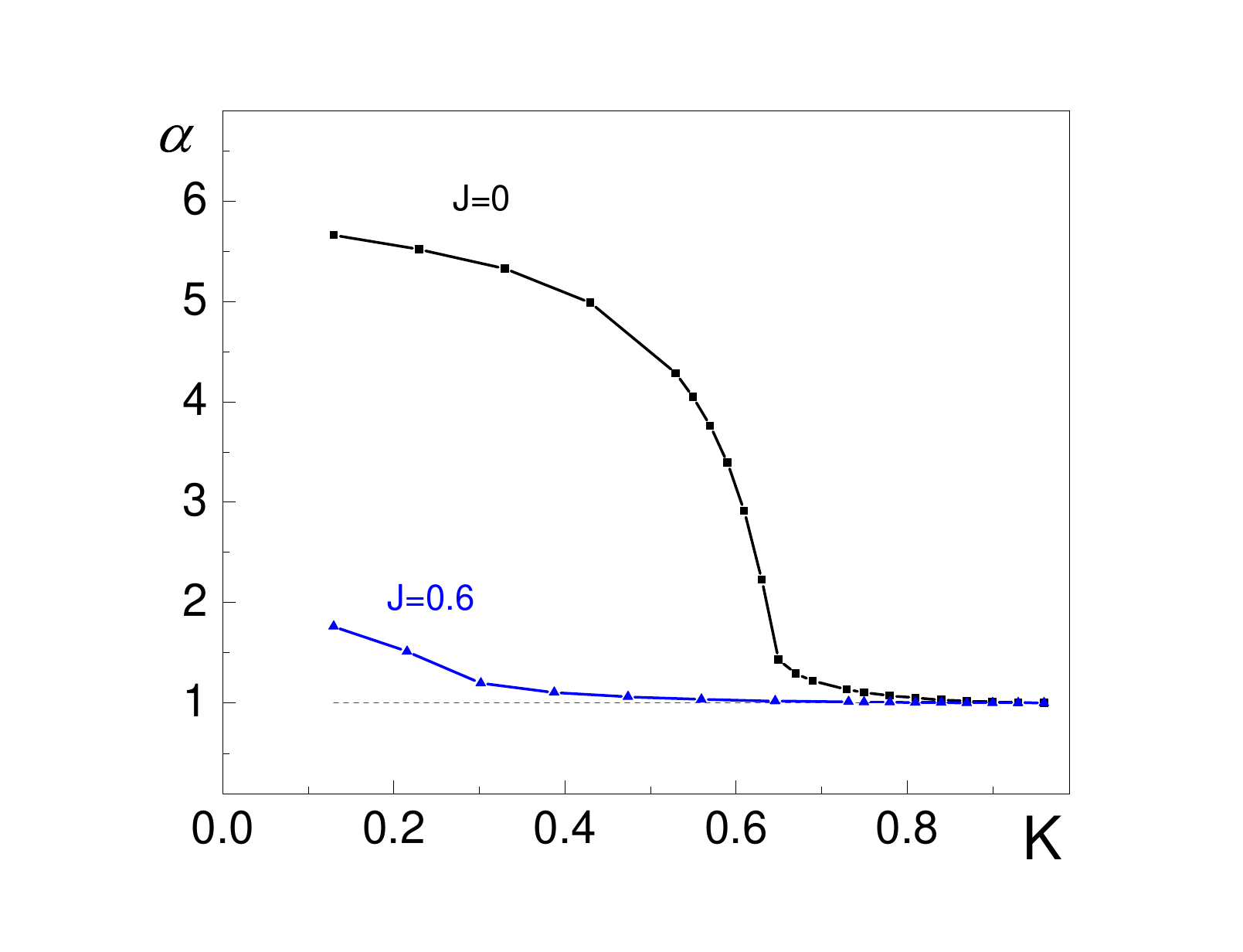}
\vspace{-0.2cm}
\caption{The power $\alpha$ in the "excitation" energy $\epsilon_1 =- \ln K^\alpha$ for two values of $J$. A remarkable feature is the approach to the limit $\alpha=1$ for $K>0.7$.}
\label{Fig:alpha}
\vspace{-0.5cm}
\end{figure}
\begin{figure}[t]
\includegraphics[width= \columnwidth]{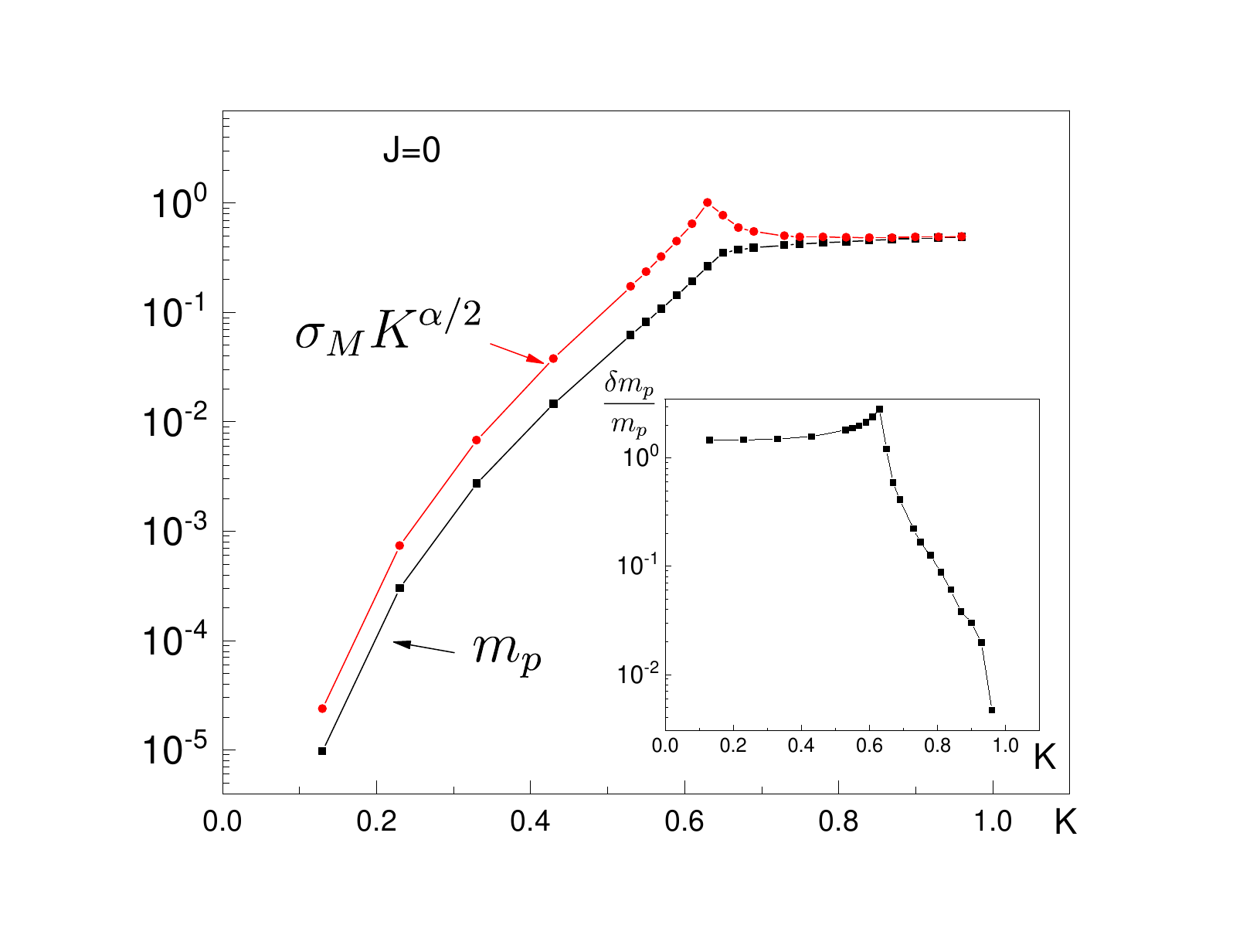}
\vspace{-0.2cm}
\caption{The mean plaquette charge $m_p$ and its fluctuation $\sigma_M$. The curves labelled as $m_p$ and $\sigma_M K^{\alpha/2}$ must coincide with each other in the case of the Poisson statistics. Inset: the fractional difference $\delta m_p /m_p = (\sigma_M K^{\alpha/2} - m_p)/m_p$. The convergence to the Poisson statistics with $\alpha \approx 1$  is reached for $K\geq 0.84$ within better than 3\%. }
\label{Fig:Mpois}
\vspace{-0.5cm}
\end{figure}
\begin{figure}[t]
\includegraphics[width= \columnwidth]{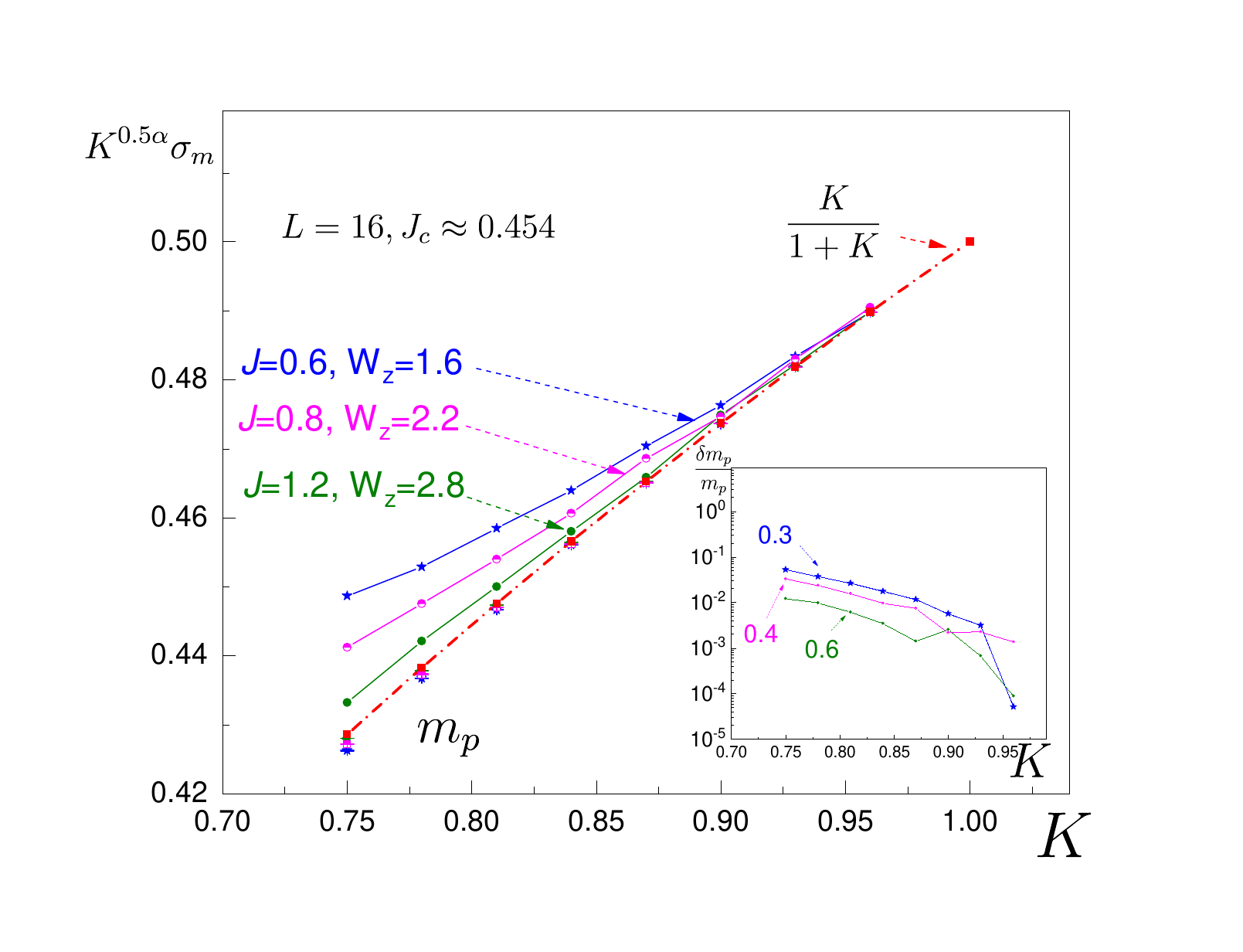}
\vspace{-0.2cm}
\caption{The mean plaquette charge $m_p$ and its fluctuation $\sigma_M$ for several values of $J$. The dash-dotted line represents the Poisson statistics for $m_p$. Symbols are the data points for $m_p$ for the corresponding values of $J$. The values of the typical values of the windings  $W_z$ for the corresponding $J$ indicate that each sample contains 1-3 large loops of the link currents.
Inset: the fractional difference $\delta m_p /m_p = (\sigma_M K^{\alpha/2} - m_p)/m_p$ for the same values of $J$ (shown close to each data set) as in the main panel. }
\label{Fig:Mpois2}
\vspace{-0.5cm}
\end{figure}

\section{Typical configurations.} 
In our simulations of the model in dual variables the key finding is that the total density (per plaquette) $m_p=\langle M_p \rangle$ of the charges $M_p$ increases with increasing $K$ and finally reaches the limiting value $m_p=1/2$ for $K=1$ ($\kappa=\infty$) with its 
fluctuation $\sigma_M =\sqrt{\langle M_p^2 \rangle - \langle M_p \rangle^2} =1/4$. In other words, asymptotic behavior of the ensemble of the plaquette charges exhibits Poisson statistics with the parity constraint becoming irrelevant. Furthermore, such a limit is realized even at finite $K$ in the phase II. This can be observed in the behavior of the "excitation" energy as $-\alpha \ln K$, with $\alpha$ crossing over from $\alpha \approx 6$  at small $K$ (in the gas phase of the "charged" cubes) to $\alpha \approx 1$ for $K>0.7$ with the deviation at the level of a few percents. In the Poisson statistics the mean value $m_p$  and its fluctuation obey $m_p=K^\alpha/(1+K^\alpha)$ and $\sigma_p =K^{-\alpha/2} m_p$, respectively. Accordingly, the quantities $m_p$ and $K^{\alpha/2}\sigma_p$ coincide. The value of $\alpha$ extracted from fitting $m_p$ by this formula gives $\alpha$ shown in Fig.~\ref{Fig:alpha}.
While deviating significantly at small $K$ (by about 200\%), the reduced fluctuation $K^{\alpha/2}\sigma_M$ approaches $m_p$ within better than 15\% as $K$ exceeds $K=0.75$. This implies the emergence of the Poisson statistics in the asymptotic limit $K\to 1$.

As can be seen in Fig.~\ref{Fig:Mpois2},  the asymptotic freedom from the parity constraint (that is, $\alpha \to 1$ and $K^{\alpha/2}\sigma_M \to m_p$) emerges starting from smaller $K$ as $J$ increases. This observation allows finding the effective free energy (per plaquette) of the ensemble as $F_b \to - \ln(1+K)$ for $K\to 1$. In terms of the deviation $\delta =1-K \approx 2\exp(-2\kappa) \to 0$, $F_b(K) - F_b(1) \to \exp(-2\kappa) $. In other words, the fluctuations between $M_p=0,1$ occur almost independently in each plaquette. 
It is important to realize that such fluctuations cannot occur inside a macroscopic loop with odd link currents (see the right panel in Fig.~\ref{Fig:Mgas2} ) because the parity requirement (\ref{Bconst}) guarantees the uniformity of the charges $M_p$ of the membrane---either $M_p=1$ (and $M_p=0$ outside) or $M_p=0$ (and $M_p=1$ outside). 
In other words, each plaquette of the membrane loses entropy if compared with a plaquette which does not belong to the membrane. 
A membrane plaquette free energy $F_M=-\ln K \approx 2\exp(-2\kappa) \to 0$.
Then, the difference $F_b - F_M \approx -\exp(-2\kappa) <0$. This implies that the total work required to create the membrane is positive as  $E_M\approx \exp(-2\kappa) L^2 >0  $, which adds an additional factor 2 to the correlation length $\xi$ in Eq.(\ref{xi}), that is, $\xi \approx 2 \ln J/ |\ln K| \sim \exp(2\kappa) \to \infty$ as $\kappa \to \infty$.

It should be mentioned that smaller odd loops with $M_p=0$ inside can decorate the membrane. In order to estimate the impact of this effect we have measured the mean density of the odd link currents per link. In the Villain version of the system this density is below 10-15\% slightly above the transition into the phase III (when the windings are of the order of unity while the stiffness is finite in the TD limit). Thus, the accuracy of our estimate of the confinement length which does not take into account the decoration effect can be set at the level of 90-85\%.  

\end{document}